\documentclass [11pt]{article}

\usepackage{graphicx}
\usepackage{amsmath,amssymb} 
\usepackage{latexsym}
\usepackage{mathrsfs}
\usepackage{bbold}
\usepackage{color}
\definecolor{red}{rgb}{1,0,0}

\makeatletter
\def\section{\@startsection {section}{1}{\z@}{-3.5ex plus -1ex minus
 -.2ex}{2.3ex plus .2ex}{\large\bf}}
\def\subsection{\@startsection{subsection}{2}{\z@}{-3.25ex plus -1ex
minus -.2ex}{1.5ex plus .2ex}{\normalsize\bf}}
\makeatother
\makeatletter

\@addtoreset{equation}{section}

\makeatother

\def\bea{\begin{eqnarray}} \def\eea{\end{eqnarray}}
\def\be{\begin{equation}} \def\ee{\end{equation}} \def\nn{\nonumber}
  \def\Z{{\bf Z}}

\newcommand{\promille}{%
  \relax\ifmmode\promillezeichen
        \else\leavevmode\(\mathsurround=0pt\promillezeichen\)\fi}
\newcommand{\promillezeichen}{%
  \kern-.05em%
  \raise.5ex\hbox{\the\scriptfont0 0}%
  \kern-.15em/\kern-.15em%
  \lower.25ex\hbox{\the\scriptfont0 00}}

\newcommand{\ol}{\overline}
\newcommand{\wt}{\tilde}

\newcommand{\GR}{\mbox{\tiny $GR$ \normalsize}}

\begin{document}

\thispagestyle{empty}

\begin{center}

\hfill SISSA-55/2011/EP \\
\hfill DFPD-2011/TH/16\\

\begin{center}

\vspace*{0.5cm}

{\Large\bf General Lepton Mixing in Holographic Composite Higgs Models}

\end{center}

\vspace{1.4cm}

{\bf Claudia Hagedorn$^{a}$ and Marco
Serone$^{b,c}$}\\

\vspace{1.2cm}

${}^a\!\!$
{\em Dipartimento di Fisica `G.~Galilei', Universit\`a di Padova
\\
INFN, Sezione di Padova, Via Marzolo~8, I-35131 Padua, Italy}

\vspace{.3cm}

${}^b\!\!$
{\em SISSA and INFN, Via Bonomea 265, I-34136 Trieste, Italy} 

\vspace{.3cm}

${}^c\!\!$
{\em ICTP, Strada Costiera 11, I-34151 Trieste, Italy}

\end{center}

\vspace{0.8cm}

\centerline{\bf Abstract}
\vspace{2 mm}
\begin{quote}

We introduce a scenario of lepton mixing in holographic composite Higgs models based on non-abelian discrete symmetries of the form
$G_f=X\times \Z_N$, broken to $\Z_2\times \Z_2\times \Z_N$ in the elementary sector and to $\Z_N^{(D)}$ in the composite sector
with $\Z_N^{(D)}$ being the diagonal subgroup of a $\Z_N \subset X$ and the external $\Z_N$.
By choosing $X = \Delta(96)$ or $\Delta(384)$, a non-vanishing  $\theta_{13}$ of order 0.1 is naturally obtained.
We apply our considerations to a 5D model in warped space for the particular cases of  
 $X = S_4, A_5, \Delta(96)$ and $\Delta(384)$ and $N=3$ or 5.
Lepton flavour violating processes and electric dipole moments  are well below the current bounds, with the exception
of $\mu\rightarrow e \gamma$ that puts a very mild constraint on the parameter space of the model, 
for all presented choices of $G_f$.

\end{quote}

\vfill

\newpage

\section{Introduction}

The data accumulated in neutrino experiments over the past years clearly show that lepton and quark mixing are vastly different.   
Several successful explanations for a lepton mixing pattern with two large angles and a small one in terms of a flavour symmetry can be found in the literature.
The most prominent pattern is tri-bimaximal (TB) mixing \cite{HPS} which can be elegantly derived with
the help of the symmetries $A_4$ \cite{A4TB} and $S_4$ \cite{S4TB}. 
Recently, the T2K \cite{T2K}  and MINOS \cite{MINOS} Collaborations published indication that the lepton mixing angle $\theta_{13}$ is non-zero. According to global fits \cite{fogli,schwetz,maltoni}, the best
fit value of $\theta_{13}$ is around $0.1 \div 0.2$ and its value is different from zero at the $3 \sigma$ level  \cite{fogli,schwetz}.  In the light of this, many of the models predicting TB mixing become disfavoured, because the
deviation from $\theta_{13}=0$ necessary to accommodate the best fit value of $\theta_{13} \sim 0.1 \div 0.2$ is too large to be explained by sub-leading corrections. 
Therefore,  new lepton mixing patterns with non-vanishing $\theta_{13}$ based on discrete non-abelian symmetries  have recently been put forward \cite{dATFH11}. The key assumptions \cite{dATFH11} (see also \cite{S4TB}) are that the neutrino and the charged lepton mass matrices are invariant  under two distinct subgroups $G_\nu$ and $G_e$ of a flavour group $G_f$, respectively, and that left-handed (LH) leptons are in an irreducible
triplet representation of $G_f$.
Non-trivial  lepton mixing is determined through the relative embedding of $G_\nu$ and $G_e$ into $G_f$. In contrast, lepton masses remain unconstrained.

The symmetry breaking pattern of $G_f$ as proposed in \cite{S4TB,dATFH11} is naturally realized in Composite Higgs Models (CHM), where the group $G_f$ is broken to $G_\nu$ in the elementary  and to $G_e$ in the composite sector.\footnote{Models in warped space making use of the discrete symmetry $A_4$ can be found in  \cite{Csaki}.}
In fact, a concrete realization of this scenario was already introduced in CHM in \cite{S4HCHM}, for the particular case $G_f = S_4\times \Z_3$, leading to TB mixing and thus $\theta_{13}=0$.

Aim of this paper is to generalize the scenario \cite{S4HCHM} in order to include lepton mixing patterns that, like in \cite{dATFH11}, lead to non-vanishing $\theta_{13}$
of order $0.1\div 0.2$. More precisely, we consider a set-up in which the discrete flavour group is $G_f= X\times \Z_N$, with $X$ being a 
non-abelian group. The additional cyclic symmetry $\Z_N$ is in general needed to keep the natural explanation of the fermion mass hierarchy
given by the Holographic CHM (HCHM) or their five-dimensional (5D) realizations. The pattern of flavour symmetry breaking is driven by symmetry considerations only,
and no  specific sources of flavour breaking are introduced.  
We focus on a scenario in which the Standard Model (SM) neutrinos are Majorana fermions and the type I see-saw mechanism 
explains the smallness of their masses. However, a similar analysis also applies to a scenario, introduced in \cite{S4HCHM}, in which neutrinos are Dirac fermions.

We choose as remnant symmetry $G_\nu$ in the elementary  sector $\Z_2 \times \Z_2 \times \Z_N$, while  $G_e$ of the composite sector is taken to be $\Z_N^{(D)}$, 
the diagonal subgroup of a $\Z_N \subset X$ and the external $\Z_N$.
As discussed already in \cite{S4HCHM}, a large breaking of $G_f$ in the composite sector  is favoured for charged leptons, because in this way large deviations from the SM $Z\tau\bar\tau $ coupling are 
suppressed. At the same time, the breaking of $G_f$  in the composite sector affecting neutrinos is required to be small, in order to not perturb too much the lepton mixing pattern determined by the choice of 
$G_f$, $G_e$ and $G_\nu$.

After a general presentation of our set-up in terms of 4D HCHM, we study the 5D models introduced in \cite{S4HCHM}  in more detail 
for the choices $(X,\Z_N)=(S_4,\Z_3)$, $(A_5,\Z_5)$, $(\Delta(96),\Z_3)$ and $(\Delta(384),\Z_3)$. The full discrete symmetry of the 5D models is actually  $G_f \times Y$, where $Y=\Z_3^\prime\times \Z_3^{\prime\prime}$
is a flavour-independent factor useful to minimize the number of allowed terms.
Keeping the prediction of the solar mixing angle $\theta_{12}$ within the experimentally allowed $3\sigma$ range requires
that the flavour symmetry breaking at the IR brane for neutrinos should be smaller than $10\%$ in all models, unless a $\Z_2$ exchange 
symmetry is imposed on the IR brane, in which case no constraint occurs.
The corrections to the mixing angles $\theta_{23}$ and $\theta_{13}$ are generically smaller. In the case of $X=\Delta(96)$ they help
 to improve the accordance of the predicted and the measured value of $\theta_{23}$. 
In the cases of  $X=S_4$ and $X=A_5$, in which the unperturbed value of $\theta_{13}$ vanishes, such corrections are not enough
to generate a value of $\theta_{13}$ of order $0.1 \div 0.2$, as favoured by the latest experimental data and global fit analyses. Overall, the patterns derived with $X=\Delta(384)$
describe the data in the best way.
Although the neutrino mass spectrum is not predicted, a normally ordered spectrum
is preferred in the 5D models, because corrections to the solar mixing angle are under much better control in this case.

For all choices of $G_f$ which we discuss, most of the Lepton Flavour Violating (LFV) processes for charged leptons are below the current experimental bounds for masses of the first Kaluza-Klein (KK) gauge resonances of order $3.5$ TeV, roughly the lowest scale allowed by electroweak considerations. The main source of such processes are Boundary Kinetic Terms (BKT) for fermions at the UV brane.
The most important bound comes from the radiative decay $\mu\rightarrow e \gamma$, $BR(\mu\rightarrow e \gamma)< 2.4\times 10^{-12}$ \cite{MEG}, 
and is passed in most of the parameter space, while the expected future bound constrains our model.
We also argue that Electric Dipole Moments (EDMs) for charged leptons are negligibly small.

The structure of the paper is as follows. In section 2 we generalize the set-up of \cite{S4HCHM} to include generic lepton mixing patterns arising from a non-trivial breaking of a flavour symmetry $G_f$.
In section 3 we apply our considerations to the 5D Majorana model of \cite{S4HCHM} and discuss the results for lepton mixing as well as constraints coming from charged LFV decays and  lepton EDMs.
We conclude in section 4. The relevant group theory of $S_4$, $A_5$, $\Delta(96)$ and $\Delta(384)$ and an explicit choice of basis for their generators can be found in the appendix.


\section{General Set-up}
\label{sec:scenario}

The set-up we consider is closely related to the one introduced in \cite{S4HCHM} for the particular choice of the discrete group $S_4\times \Z_3$.
We mainly emphasize here the key differences with respect to \cite{S4HCHM}, referring the reader to \cite{S4HCHM} for further details.
We consider in this paper only CHM with Majorana neutrinos, since they overall seem to perform better than the CHM with Dirac neutrinos, but similar considerations
apply to the latter case as well.

The Lagrangian of the CHM consists of an elementary, a composite and a mixing sector \cite{Agashe:2004rs}: 
\be
{\cal L}_{tot} = {\cal L}_{el} + {\cal L}_{comp}+{\cal L}_{mix}\,.
\label{Ltot}
\ee
We assume that ${\cal L}_{mix}$ is invariant under the discrete flavour symmetry $G_f=X \times \Z_N$, with $X$ a non-abelian group\footnote{The group $X$ can in principle be infinite, but all examples we present in the following make use of a finite $X$.} which has the following 
features, see \cite{S4TB,dATFH11}: i) it contains, at least, one (faithful) irreducible three-dimensional representation {\bf 3}, as which LH leptons and right-handed (RH) neutrinos transform,\footnote{We assume for simplicity that LH leptons and RH neutrinos transform in the same way under $G_f$.} and ii)  it contains $\Z_2\times \Z_2$ and $\Z_N$ as non-commuting subgroups.
The former requirement ensures the possibility to determine all mixing angles through the choice of $G_f$, $G_e$ and $G_\nu$, while the latter ensures that the resulting mixing pattern is non-trivial,
because lepton mixing corresponds to the mismatch in the embedding of the two subgroups $\Z_2\times \Z_2$ and $\Z_N$ into $X$.

The symmetry $G_f$ is broken in the elementary sector  to $G_\nu=\Z_2\times \Z_2\times \Z_N$, where $\Z_2\times \Z_2 \subset X$, and
in the composite sector to $G_e=\Z_N^{(D)}$, the diagonal subgroup of the external $\Z_N$ and $\Z_N\subset X$. Thus,  all terms of ${\cal L}_{el}$ are invariant
under $G_\nu$, while all terms of ${\cal L}_{comp}$ under $G_e$. 

$G_e$ is chosen as $\Z_N^{(D)}$ with $N\geq 3$ in order to distinguish the three generations of charged leptons and consequently to explain the observed hierarchy among their masses. More precisely, it has to be the diagonal subgroup of $\Z_N\subset X$ and the external $\Z_N$
because LH leptons have to be assigned to a {\bf 3} of $X$ for lepton mixing,  
while RH charged leptons transform trivially under $X$ and carry only non-trivial charge under the external $\Z_N$.
Obviously, in order to distinguish among the three generations, their charges have to be different.
We consider the group $\Z_N$ for simplicity. In a more general set-up this group can be replaced by a product of cyclic symmetries, such as $\Z_2 \times \Z_2$.
If the group $X$ has three or more inequivalent one-dimensional representations, the additional cyclic group factor might be abandoned, because it is then possible
to distinguish the three generations of (RH) charged leptons with the help of $X$ alone.
   
$G_\nu$ consists of a Klein group $\Z_2\times \Z_2$ and of the external $\Z_N$. The Klein group is the maximal symmetry preserved by a Majorana mass matrix  in the case of three neutrinos \cite{S4TB},
and at the same time it can guarantee the existence of three independent parameters corresponding to the neutrino masses.\footnote{In the case of Dirac neutrinos $G_\nu$ is not constrained to contain a Klein group, since the symmetry preserved by a Dirac neutrino mass matrix can be any product of cyclic symmetries
which allows to distinguish among the three generations. The simplest case is then that $G_\nu$ is a product of a cyclic symmetry $\Z_M$ and the external $\Z_N$ with $M,N \geq 3$.}
The external $\Z_N$ does not play any direct role for the generation of lepton mixing, because neither LH leptons nor RH neutrinos transform under it, but it automatically keeps the kinetic terms of RH charged leptons flavour diagonal in the elementary sector (see below).\footnote{In a set-up without the external $\Z_N$ the RH charged leptons should transform as three distinct singlets under the subgroup $\Z_2 \times
\Z_2$ contained in $G_\nu$ (and thus under $X$) in order to keep their kinetic terms flavour diagonal in the elementary sector.}

In the basis in which the generator $G_N$ of $\Z_N\subset X$  is diagonal for {\bf 3},
\be
G_N = \left(\begin{array}{ccc}
\omega_N^{n_e} & 0 & 0\\
0 & \omega_N^{n_\mu} & 0\\ 
0 & 0 & \omega_N^{n_\tau} 
\end{array}
\right)\,, \;\;\; \mbox{with} \;\;\;   n_e \neq n_\mu \neq n_\tau \, ,\;\; \omega_N = e^{2\pi i/N} \, ,
\label{GeDiag}
\ee
 the generators $G_1$ and $G_2$ of $\Z_2\times \Z_2$  are of the form
\be
G_1 = V G_1^{diag} V^\dagger, \ \ \ \ \ G_2 = V G_2^{diag} V^\dagger\,,
\label{g1g2V}
\ee
with
\be
\label{g1g2diag}
G_1^{diag} = \left( \begin{array}{ccc}
 1 & 0 & 0\\
0& -1 &0\\
0&0 & -1
\end{array}
\right)\,, \ \ \ \ G_2^{diag} = \left( \begin{array}{ccc}
 -1 & 0 & 0\\
0& 1 &0\\
0&0 & -1
\end{array}
\right)\,,
\ee
and $V$ a unitary matrix.

The elementary sector contains three generations of SM LH and RH leptons $l_L^\alpha$, $l_R^\alpha$ and three RH neutrinos $\nu_R^\alpha$.  The LH leptons $l_L^\alpha$ and the RH neutrinos $\nu_R^\alpha$ transform as $({\bf 3},1)$ under $X\times \Z_N$,
while the RH charged leptons $l_R^\alpha$ transform as
$({\bf 1},\omega_N^{n_\alpha})$.  
The elementary Lagrangian is 
\be
{\cal L}_{el} = \bar l_L^\alpha i  \slash \!\!\!\!D l_L^\alpha + \bar l_R^\alpha i  \slash \!\!\!\!D l_R^\alpha + \bar \nu_R^\alpha i  \slash \!\!\!\partial \nu_R^\alpha - \frac 12  (\ol{\nu^{c}_R}^\alpha M_{\alpha\beta} \nu_R^\beta +h.c.)\,,
\label{Lele}
\ee
where $M$ is the most general mass matrix invariant under $\Z_2\times \Z_2 \times \Z_N$, of the form 
\be
M = V^\star M_D V^{\dagger} \,,
\label{MV}
\ee
with $V$ as in (\ref{g1g2V})  and $M_D$ a diagonal matrix containing three independent complex parameters.

The composite sector is an unspecified strongly coupled theory, that gives rise, among other states, to a composite
SM Higgs field and vector-like fermion resonances $\Psi$ mixing with the SM fields.
The fermion mixing Lagrangian ${\cal L}_{mix}$ is
\be
{\cal L}_{mix}  =  \frac{\lambda_{l_L}}{\Lambda^{\gamma_{lL}}} \bar{l}_L^\alpha \Psi_{l_L,R}^\alpha + \frac{\lambda_{l_R}^\alpha}{\Lambda^{\gamma_{lR}^\alpha}} \bar{ l}_R^\alpha \Psi_{l_R,L}^\alpha +\frac{\lambda_{\nu_R}}{\Lambda^{\gamma_{\nu R}}} \bar{\nu}_R^\alpha\Psi_{\nu_R,L}^\alpha+h.c. 
\label{Lmix}
\ee
where $\Lambda$ is the UV cut-off scale of the composite sector and $\Psi_{l_L}^\alpha$, $\Psi_{l_R}^\alpha$ and $\Psi_{\nu_R}^\alpha$ are fermion resonances  transforming under $G_f$ in the same way as $l_L^\alpha$, $l_R^\alpha$ and $\nu_R^\alpha$, respectively.
The mixing parameters $\lambda_{l_L}$ and $\lambda_{\nu_R}$ 
are flavour universal, because two triplets of $X$ are coupled to each other, while $\lambda_{l_R}^\alpha$ are flavour diagonal, but non-universal, since RH charged leptons and $\Psi_{l_R}^\alpha$  are singlets under $X$. Integrating out the composite sector gives rise to
 the following charged lepton mass matrix (in left-right convention, $\bar\psi_L M_l \psi_R$):
\be
M_{l,\alpha\beta} \sim 
   b_\alpha v_H\lambda_{l_L}\lambda_{l_R}^\alpha\delta_{\alpha\beta}\Big(\frac{\mu}{\Lambda}\Big)^{\gamma_{lR}^\alpha+\gamma_{lL}}\,,  \label{mlept4D}  
\ee
where $v_H\simeq 250$ GeV, $\mu$ is the ${\cal O}$(TeV) scale at which the composite theory becomes strongly coupled and $b_\alpha$ are ${\cal O}(1)$ coefficients. 
In this basis the charged lepton mass matrix is flavour diagonal and non-trivial mixing is encoded in the light neutrino mass matrix.
The latter arises upon integrating out the RH neutrinos $\nu^\alpha_R$ 
\be
M_{\nu,\alpha\beta} \simeq \hat b_\alpha \hat b_\beta  v_H^2 \lambda_{l_L}^2 \lambda_{\nu_R}^2 \Big(\frac{\mu}{\Lambda}\Big)^{2(\gamma_{\nu R}+\gamma_{l L})}
  M^{-1}_{\alpha\beta} 
\simeq \hat b^2 v_H^2 \lambda_{l_L}^2 \lambda_{\nu_R}^2 \Big(\frac{\mu}{\Lambda}\Big)^{2(\gamma_{\nu R}+\gamma_{l L})}
 \Big( V M_D^{-1} V^T\Big)_{\alpha\beta} 
\label{mMajo}
\ee
with $\hat b_\alpha$ being order one coefficients.
The second relation holds in the limit of universal $\hat b_\alpha$.  Only in this limit the lepton mixing matrix  $U_{PMNS}$ is given by
\be
U_{PMNS} = V \, .
\label{VUPMNS}
\ee
Obviously, a sensible choice of $V$ implies that the resulting mixing angles are in good agreement with the experimental data.
 Deviations from the universality of $\hat b_\alpha$ lead to corrections of the
lepton mixing angles and, in order to keep the accordance with experimental data, $\hat b_\alpha$ generically have to be universal at a level of $\lesssim 10 \%$ (depending slightly on the choice of $V$).
This condition is equivalent to requiring that the breaking of $G_f$ in the composite sector should be small for neutrinos. On the other hand, it can be of order one for charged leptons, because
the values of $b_\alpha$ do not have a direct impact on the lepton mixing.

The remnant symmetry $\Z_N^{(D)}$ renders all couplings flavour diagonal in the composite sector and flavour violation is only present in (\ref{Lele}).
All flavour changing processes are then negligibly small, since they are suppressed
by the large Majorana mass of the RH neutrinos. The main source of flavour violation arises from the elementary sector, if the most general kinetic terms of the SM fermions 
compatible with the flavour symmetries are taken into account. These are  of the form
\be
 \bar l_L (1+Z_l)  i  \slash \!\!\!\!D l_L + \bar l_R (1+\tilde Z_l^D)  i  \slash \!\!\!\!D l_R 
+ \bar \nu_R (1+Z_\nu)  i  \slash \!\!\!\partial \nu_R 
  \label{Z4Dgen}
\ee
with $Z_l = V Z_l^D V^\dagger$, $Z_\nu = V Z_\nu^D V^\dagger$ and $Z_l^D$, $\tilde Z_l^D$ and $Z_\nu^D$ diagonal matrices. As explained above, the $\Z_N$ contained in $G_\nu$
forbids  flavour violating kinetic terms for $l^\alpha_R$.
Non-trivial LFV processes (and further corrections to the lepton mixing (\ref{VUPMNS})) are now generated and are proportional to the non-diagonal entries of the matrices $Z_l$ and $Z_\nu$ in (\ref{Z4Dgen}). 
In the limit in which the composite sector is $G_f$-invariant, one can go to a basis in which
the whole Lagrangian ${\cal L}_{tot}$ in (\ref{Ltot}) is flavour diagonal, since in this limit all couplings and mass terms of fermion resonances $\Psi$ forming triplets under $G_f$ are flavour universal. Thus, the actual amount of LFV is controlled by the size of the flavour violation in the elementary {\it and} the flavour non-universality in the composite sector
and is consequently suppressed with respect to an anarchic scenario with no flavour symmetries. 

Without further constraints on the breaking of $G_f$ in the elementary sector,
the elements of $Z^D_{l,\nu}$ are expected to be uncorrelated ${\cal O}(1)$ parameters, possibly leading to too large flavour violating processes and corrections to the lepton mixing. The actual effect induced by $Z_{l,\nu}$, however, depends on the degree of compositeness of $l_L^\alpha$ and $\nu_R^\alpha$ 
(in turn determined by  the mass mixing terms $\lambda_{l_L}$ and $\lambda_{\nu_R}$), 
since the kinetic terms of the elementary fields always receive a contribution coming from the strongly coupled sector,
when the fermion resonances $\Psi$ are integrated out (c.f. (2.7) of \cite{S4HCHM}). 
Thus we have to rescale the fields to canonically normalize their kinetic terms,  $\psi \rightarrow \psi/\sqrt{A_\psi }$. If the contribution from the composite sector dominates, $A_\psi \gg 1$, 
the rescaled parameters $Z_{l,\nu}/A_{l,\nu}$ become suppressed.  
As explained in \cite{S4HCHM}, $\lambda_{\nu_R}$ is a relevant coupling and $\nu_R^\alpha$ 
are mostly composite fields. The effect 
of $Z_\nu$ is thus negligible, being suppressed by the large value of $A_\nu$. 
The mixing term $\lambda_{l_L}$, on the other hand, should be irrelevant, otherwise too large deviations from the SM gauge couplings of $l_L^\alpha$  would occur (see (\ref{deltagtau}) below), but it can be very close to be marginal, $\gamma_{lL}\simeq 0$.
In this case the kinetic term of $l_L^\alpha$ can still be dominated by the contribution coming from the composite sector due to a large logarithmic running from the scale $\Lambda$ to $\mu$, 
 and again the rescaled parameters $Z_l/A_l$ turn out to be small (compare (\ref{BKTeff2})).
Summarizing, the flavour violating effects and the corrections to lepton mixing induced by $Z_{l,\nu}$ are naturally suppressed by the dynamics in the composite sector.

Flavour symmetries can be important also for flavour conserving observables, such as the EDMs.
Using standard conventions,  we denote the charged lepton EDMs $d_\alpha$ as the coefficient of the dimension five operator $(-i/2) \bar l_\alpha \sigma^{\mu\nu}\gamma_5 l_\alpha F_{\mu\nu}$. 
Being the latter a CP-odd operator, EDMs can be generated, if CP violation is present in the lepton sector.
In absence of flavour symmetries, a rough one-loop estimate, assuming generic masses and complex couplings in the composite sector,  gives
\be
d_\alpha \sim \frac{eM_{l,\alpha}}{16\pi^2}  \frac{Y^2}{m_\Psi^2}\,,
\label{dledm}
\ee
where $M_{l,\alpha}$ are the charged lepton masses, 
$Y$ represents a typical ${\cal O}(1)$ Yukawa coupling involving the composite fermion resonances and $m_\Psi$ their typical mass.
The appearance of $M_{l,\alpha}$ in (\ref{dledm}) is a consequence of the partial compositeness of the SM fermions which implies that the EDMs vanish unless both mass mixing
for LH and RH charged leptons are inserted in the one-loop diagram. Despite the explicit dependence on the charged lepton mass, the strongest bound comes from the EDM of the electron, 
$|d_e|\lesssim  10^{-27} e$ cm \cite{Hudson:2011zz}, giving  $m_\Psi \gtrsim  10 \,Y$ TeV. 
In our scenario, however,  the leading order result (\ref{dledm}) vanishes for canonical kinetic terms (\ref{Lele}), because the relevant non-trivial phases can be removed through field redefinitions.\footnote{This is not true in general, if one considers non-minimal scenarios, in which for example SM fermions mix with 
more than one state of the composite sector.}
For general kinetic terms (\ref{Z4Dgen}) and flavour symmetry breaking in the composite sector,
the leading order term no longer vanishes and is estimated to be, assuming again arbitrary complex couplings in the composite sector,
\be
d_{\alpha} \sim \frac{eM_{l,\alpha}}{16\pi^2}  \frac{Y^2}{m_\Psi^2}\frac{\delta m_\Psi}{m_\Psi} Z_l^2\,,
\label{dledm2}
\ee
where $\delta m_\Psi$ is the inter-generational mass splitting of the fermion resonances. The last two terms in (\ref{dledm2}) are essential suppression factors that can significantly reduce
the size of $d_\alpha$.

Let us conclude this section by showing how deviations from the SM $Zl_{\alpha}\bar l_{\alpha} $ couplings disfavour a small breaking of the flavour symmetry in the composite sector for charged leptons, $b_\alpha\ll1$ in (\ref{mlept4D}). For simplicity we set $Z_l=\tilde Z_l^D=Z_\nu=0$, their effect being sub-leading.
The coupling deviations arise from mixing of the SM leptons with fermion resonances, induced by the mixing terms in (\ref{Lmix}).
One schematically has
\be
\frac{\delta g_{l_{\alpha i}}}{g_{l_{\alpha i}}} \sim \frac{v_H^2}{m_\Psi^2} (\lambda_{l_i}^{\alpha})^2 \Big(\frac{\mu}{\Lambda}\Big)^{2\gamma_{li}^\alpha}, \ \ \ i = L,R\,.
\label{deltagtau}
\ee
Using the charged lepton mass formula (\ref{mlept4D}), we can write
\be
\frac{\delta g_{l_{\alpha L}}}{g_{l_{\alpha L}}} \frac{\delta g_{l_{\alpha R}}}{g_{l_{\alpha R}}}  \sim \frac{M^2_{l,\alpha}}{m^2_\Psi} \frac{v^2_H}{m^2_\Psi} \frac{1}{b_\alpha^2}\,.
\label{deltagtau2}
\ee
The tension between having parametrically small $b_\alpha$ and sufficiently small gauge coupling deviations is obvious from (\ref{deltagtau2}).
This is particularly important for the tau lepton due to its larger mass.
Deviations from the  SM gauge couplings for (charged and neutral) leptons have been constrained by LEP at the per mille level \cite{LEPZcoupling}. For fermion resonances at the TeV scale, 
we see that the right-hand side of (\ref{deltagtau2}) is below $10^{-6}$ for $b_\tau$ slightly below one. 
Notice that one might actually cancel the leading term (\ref {deltagtau}) in $\delta g_{l_{\alpha L}}$ for either charged leptons or neutrinos using appropriate symmetries \cite{Agashe:2006at}, 
but not both at the same time. The relation (\ref{deltagtau2}) is then always valid for at least one of the two components of the LH doublet $l_L^\alpha$.

\section{5D Realizations}
\label{sec:leptonM}

Models with gauge-Higgs unification in warped space based on an $SO(5)\times U(1)_X$ gauge symmetry \cite{Agashe:2004rs,GHUws} are
an explicit and particularly interesting weakly coupled description of the scenario outlined in section \ref{sec:scenario}.\footnote{Notice that warping is not a necessary ingredient. For viable
models in flat space see \cite{Panico:2010is}.}
We consider flavour groups of the form:
\be
G_f \times Y \, .
\ee
The additional discrete symmetry $Y$,
\be
Y=\Z_3^\prime \times \Z_3^{\prime\prime} \, ,
\ee
is introduced in order to minimize the number of couplings in the bulk and at the branes.
We present four examples in the following leading to different results for the lepton mixing angles.
Our first choice of $G_f$ is 
\be
G_f  =  S_4 \times \Z_3
\ee
which is discussed in \cite{S4HCHM} and leads to TB mixing, see (\ref{VS4}). We repeat its analysis here for completeness.
The choice
\be
G_f  = A_5 \times \Z_5
\ee
gives rise to the so-called Golden Ratio (GR)
mixing in which the solar mixing angle $\theta_{12}$ is determined in terms of $\phi =\frac{1}{2}(1+\sqrt{5})$ \cite{A5FP},
while vanishing $\theta_{13}$ and maximal $\theta_{23}$ are predicted, see (\ref{VA5}).
We include its discussion, albeit it might seem to be disfavoured due to the prediction $\theta_{13}=0$, in order to show an example
in which the external $\Z_N$ factor is different from $\Z_3$.
As third and forth choice, we discuss the cases
\be
G_f  = \Delta(96) \times \Z_3\;, \;\;
G_f  =  \Delta(384)  \times \Z_3 \, ,
\ee
since it has recently been shown that these groups can naturally lead to $\theta_{13} \sim 0.1 \div 0.2$ \cite{dATFH11}.
They give rise to two inequivalent mixing patterns each (differing in the value of the angle $\theta_{23}$), and thus
we get in total four different possibilities. Following the notation of \cite{dATFH11}, we call them {\tt M1} and {\tt  M2} for $\Delta(96)$, see (\ref{VD96M1}) and (\ref{VD96M2}),  
and {\tt M3} and {\tt M4} for $\Delta(384)$, see (\ref{VD384M3}) and (\ref{VD384M4}), respectively.

\begin{table}
\centering
\begin{tabular}{|c|c|c|c|}\hline
& Bulk & UV & IR\\[0.05in]
\cline{2-4}
 & $G_{f} \times Y$ & $G_{f,\mathrm{UV}} \times Y_{\mathrm{UV}}$ & \rule[0.1in]{0cm}{0cm} $G_{f,\mathrm{IR}} \times Y_{\mathrm{IR}}$\\[0.05in]
\hline
 &   & \rule[0.05in]{0cm}{0cm} $(1,-1,1,\omega_3)$ & \\
 $\xi_{l,\alpha}$         &  $({\bf 3}, 1, \omega_3,\omega_3)$                                 & $(-1,1,1,\omega_3)$ & $(\omega_N^{n_\alpha},\omega_3)$\\
          &                                   & $(-1,-1,1,\omega_3)$ & \\[0.05in]
\hline
\rule[0.1in]{0cm}{0cm}$\xi_{e,\alpha}$ &  $({\bf 1}, \omega_N^{n_ \alpha}, \omega_3, \omega_3)$ & $(1,1,\omega_N^{n_ {\alpha}},\omega_3)$ & $(\omega_N^{n_\alpha},\omega_3)$\\[0.05in]
\hline
 &   & \rule[0.05in]{0cm}{0cm} $(1,-1,1,1)$ & \\
 $\xi_{\nu,\alpha}$         &  $({\bf 3}, 1, \omega_3,1)$                                 & $(-1,1,1,1)$ & $(\omega_N^{n_\alpha},\omega_3)$\\
          &                                   & $(-1,-1,1,1)$ & \\[0.05in]
\hline
\end{tabular}
\caption{Transformation properties of the 5D multiplets $\xi_{l,\alpha}$, $\xi_{e,\alpha}$ and $\xi_{\nu,\alpha}$ under $G_{f} \times Y$ and the subgroups $G_{f,\mathrm{UV}} \times Y_{\mathrm{UV}}$ and $G_{f,\mathrm{IR}} \times Y_{\mathrm{IR}}$.
The values of $N$ and $n_\alpha$ for each group can be found in table \ref{table:groups}.}
\label{table:flavourM}
\end{table}
The flavour symmetry is broken at the UV and IR branes 
to 
\be
G_{f,\mathrm{UV}} \times Y_{\mathrm{UV}}=\Z_2\times \Z_2\times \Z_N\times \Z^{\prime\prime}_3 \,, \;\;\;\; 
G_{f,\mathrm{IR}}\times Y_{\mathrm{IR}}=\Z^{(D)}_N\times \Z^{\prime}_3 \; ,
\ee 
with $N=3$ in the case of $X=S_4$, $\Delta(96)$ and $\Delta(384)$, and $N=5$ for $A_5$.

The lepton particle content of the model is identical to the one of \cite{S4HCHM} with respect to the gauge group: three 5D bulk fermions  $\xi_{l,\alpha}$, 
$\xi_{e,\alpha}$ and  $\xi_{\nu, \alpha}$, in the fundamental, adjoint and singlet representations of $SO(5)$ are introduced (for details of the notation see \cite{S4HCHM}). 
All of them have vanishing $U(1)_X$ charge. Their flavour properties are reported in tables  
\ref{table:flavourM} and \ref{table:groups}. 

The most general $G_{f,\mathrm{IR}}\times Y_{\mathrm{IR}}$ invariant mass terms at the IR brane are
\be
-\mathcal{L}_{\mathrm{IR}} = \left(\frac{R}{R'}\right)^4 \sum \limits_{\alpha=e,\mu,\tau}\left(m_{\mathrm{IR},\alpha}^l 
\left(\ol{\wt{L}}_{1,\alpha L} \wt{L}_{2,\alpha R} + \ol{L}_{\alpha L} \hat{L}_{\alpha R}\right)
+m_{\mathrm{IR},\alpha}^\nu  \, \ol{\hat{\nu}}_{\alpha L} \nu_{\alpha R} 
+ h.c. \right)\,.
\label{IRmasses}
\end{equation}
In the particular bases chosen for the different groups $G_f$, see section \ref{sec:scenario} and the appendix, these terms are
 flavour diagonal. The fields $\wt{L}_{1,\alpha}$, $L_{\alpha}$ and $\hat \nu_\alpha$ are components of the 5D multiplet $\xi_{l,\alpha}$,
$ \wt{L}_{2,\alpha}$ and $\hat{L}_{\alpha}$ are contained in $\xi_{e,\alpha}$ and $\nu_\alpha=\xi_{\nu,\alpha}$. 
The only $G_{f,\mathrm{UV}}\times Y_{\mathrm{UV}}$ invariant mass terms at the UV brane 
are Majorana mass terms for RH neutrinos:
\be
-\mathcal{L}_{\mathrm{UV}} = \frac{1}{2} \, \ol{\nu_{\alpha R}^c} \mathcal{M}_{\mathrm{UV},\alpha \beta} \nu_{\beta R} + h.c.
\label{UVmasses}
\ee
with 
\be
\mathcal{M}_{\mathrm{UV}} = V^\star  m_{\mathrm{UV}} V^\dagger \,,
\label{MUV}
\ee
$m_{\mathrm{UV}}=\mbox{diag} \left( m_{\mathrm{UV},e}\, , \; m_{\mathrm{UV},\mu}\, , \; m_{\mathrm{UV},\tau} \right)$ and $V$ as in (\ref{g1g2V}). 
Up to (removable) phases and signs, the explicit form of $V$ is as follows:
\bea
X=S_4 &:& V=U_{TB} = \left( \begin{array}{ccc}
 \sqrt{\frac{2}{3}} & \sqrt{\frac{1}{3}} & 0\\
 -\sqrt{\frac{1}{6}} & \sqrt{\frac{1}{3}} & \sqrt{\frac{1}{2}}\\
 -\sqrt{\frac{1}{6}} & \sqrt{\frac{1}{3}} & -\sqrt{\frac{1}{2}}
\end{array}
\right)\,, \label{VS4}\\
&&\nn\\
X=A_5 &:& V=U_{GR}=\left(
\begin{array}{ccc} 
\cos \theta_{12}^{\GR}&-\sin \theta_{12}^{\GR}&0\\
\frac{\sin\theta_{12}^{\GR}}{\sqrt 2}&\frac{ \cos\theta_{12}^{\GR}}{\sqrt 2}&\frac{1}{\sqrt 2}\\
\frac{\sin\theta_{12}^{\GR}}{\sqrt 2}&\frac{ \cos\theta_{12}^{\GR}}{\sqrt 2}&-\frac{1}{\sqrt 2}
\end{array}
\right)\,,  \;\;\; \mbox{with} \;\;\; \tan\theta_{12}^{\GR}=1/\phi  \,, \label{VA5}
\eea
\begin{table}
\centering
\begin{tabular}[t]{|c|c|c|c|c|c|c|}\hline
$X$ & $S_4$ & $A_5$ & $\Delta(96)$, {\tt M1} & $\Delta(96)$, {\tt M2} & $\Delta(384)$, {\tt M3} & $\Delta(384)$, {\tt M4}  \\[0.05in]\hline 
 $N$ &   3 &   5 &  3 & 3 &  3 &  3  \\[0.05in] \hline
 $n_\alpha$ &  (0,2,1)  &  (0,1,4) &  (2,1,0) &  (2,0,1) &  (1,2,0) &  (1,0,2)  \\[0.05in]
\hline
\end{tabular}
\caption{Values of $N$ and $n_\alpha$ for the different choices of non-abelian discrete groups $X$ and mixing patterns.}
\label{table:groups}
\end{table}
\bea
X=\Delta(96), \, \mbox{{\tt M1}} &:& V=\frac{1}{\sqrt{3}}
\left(
\begin{array}{ccc}
-\frac{1}{2}(\sqrt{3}+1) & 1 & \frac{1}{2}(\sqrt{3}-1) \\
\frac{1}{2}(\sqrt{3}-1) & 1 & -\frac{1}{2}(\sqrt{3}+1) \\
   1                    & 1 & 1
\end{array}
\right) \,, \label{VD96M1}\\ &&\nn \\
 X=\Delta(96), \, \mbox{{\tt M2}} &:& V \; \mbox{equal to the one of {\tt M1} with 2$^{\rm nd}$ and 3$^{\rm rd}$ rows exchanged}\,, 
  \label{VD96M2}
 \eea
 \bea
X=\Delta(384), \, \mbox{{\tt M3}} &:& V=\frac{1}{\sqrt{3}}
\left(
\begin{array}{ccc}
-\frac{1}{2}\sqrt{4+\sqrt{2}+\sqrt{6}}&1& -\frac{1}{2}\sqrt{4-\sqrt{2}-\sqrt{6}}\\
\frac{1}{2}\sqrt{4+\sqrt{2}-\sqrt{6}}&1&-\frac{1}{2}\sqrt{4-\sqrt{2}+\sqrt{6}}\\
\sqrt{1-\frac{1}{\sqrt{2}}}&1&\sqrt{1+\frac{1}{\sqrt{2}}}
\end{array}
\right)\,,  \label{VD384M3}\\ && \nn \\
X=\Delta(384), \, \mbox{{\tt M4}} &:& V \; \mbox{equal to the one of {\tt M3} with 2$^{\rm nd}$ and 3$^{\rm rd}$ rows exchanged}\,.
\label{VD384M4}
\eea
In order to discuss the result for lepton mixing analytically, we consider charged lepton and neutrino mass matrices in the Zero Mode Approximation (ZMA), including the effect of the dominant flavour violating BKT
\be
{\cal L}_{BKT} = \bar L_L(x,R) (R \hat Z_l) i  \slash \!\!\!\!D L_L(x,R) \,,
\label{LBKT}
\ee
with $\hat Z_l$ being constrained by $\Z_2 \times \Z_2 \times \Z_N$ to be of the form $\hat Z_l = V {\rm diag} \,(\hat z_{el},\hat z_{\mu l},\hat z_{\tau l}) V^\dagger$.\footnote{The coefficients $\hat z_{el}$, $\hat z_{\mu l}$ and $\hat z_{\tau l}$ were denoted by
$z_{el}$, $z_{\mu l}$ and $z_{\tau l}$ in \cite{S4HCHM}. We use the latter notation for the entries of $Z_l$ defined in (\ref{BKTeff2}).}
The effective BKT relevant for the single KK modes are obtained by multiplying $\hat Z_l$ with the square of their wave function profile evaluated at the UV brane.
For zero modes we get,  if the bulk mass parameter $c_l$ of the fermions $\xi_{l,\alpha}$ fulfills $c_l=1/2+\delta_c$, at linear order in $\delta_c$, 
\be
Z_l \simeq \Big(\log^{-1}\frac{R^\prime}{R}+\delta_c \Big)\hat Z_l \simeq  \Big(\frac{1}{35}+ \delta_c\Big) \hat Z_l \,,
\label{BKTeff2}
\ee
taking $R$ of the order of the inverse of the reduced Planck mass and $R'$ of the order of the inverse of the TeV scale.
The matrix $\hat Z_l$ in (\ref{BKTeff2}) should roughly be identified with $Z_l$ introduced in (\ref{Z4Dgen}), and $1/35$ 
is the 5D counterpart of the suppression factor coming from the composite sector, discussed in section 2. The latter plays a crucial role in naturally suppressing most of the LFV processes well below their current experimental bounds and in keeping the corrections to the lepton mixing small.
The charged lepton mass matrix in the ZMA reads, after canonical normalization of the kinetic terms (\ref{LBKT}) and additionally rotating LH charged leptons with $V^\dagger$
(again all relevant notation can be found in \cite{S4HCHM})
\be
M_{l, \alpha\beta} =   \frac{h}{\sqrt{2}R^\prime}  \, f_{c_l} f_{-c_\beta} \left(V \frac{1}{\sqrt{1+Z_l^D}} V^{\dagger} \frac{m_{\mathrm{IR}}^l}{\sqrt{\rho}} 
 \right)_{\alpha\beta} \,. \label{CLMassMatrix} 
\ee
The light neutrino mass matrix, after integrating out the heavy RH neutrinos and canonically normalizing the kinetic terms (this time without rotating LH neutrinos with $V^\dagger$), is
\be
\label{NuMassMatrix}
M_{\nu,\alpha\beta} =  \frac{h^2}{2R^{\prime 2}} f_{c_l}^2 
\left(\frac{R'}{R} \right)^{2 c_\nu+1} \!\! \left( \frac{1}{\sqrt{1+Z_l^D}} V^{\dagger}
\frac{m_{\mathrm{IR}}^\nu}{\sqrt{\rho}} 
V\frac{R}{m_{\mathrm{UV}}} V^T \frac{m_{\mathrm{IR}}^\nu}{\sqrt{\rho}}
V^\star   \frac{1}{\sqrt{1+Z_l^D}} \right)_{\alpha\beta}\,.
\ee
In this basis the charged current is of the form 
\be
\bar l_L W^{-} V  \nu_L
\ee
  which coincides with the result given in (\ref{VUPMNS}) in the limit in which the BKT are set to zero and the mass parameters $m_{\mathrm{IR},\alpha}^\nu$ as well as the factors $\rho_\alpha$ are taken to be universal. 
  In the phenomenologically interesting region of the parameter space in which the bulk mass parameter $c_l$ of $\xi_{l,\alpha}$ is close to 1/2, the mass parameters $m_{\mathrm{IR},\alpha}^l$ can be of order one
  without affecting considerably the universality of the parameters $\rho_\alpha$ and thus  
  the results for lepton mixing. On the other hand, we still need to assume a small breaking of $G_f$ at the IR brane in the neutrino sector, i.e.
\be
 m_{\mathrm{IR},\alpha}^\nu=m_{\mathrm{IR},0}^\nu (1+\delta_\alpha) \, ,
 \label{delta_a}
 \ee
 with $|\delta_\alpha| \ll1$, in order to keep their impact on the mixing angles under control. As explained in \cite{S4HCHM}, the parameters $\hat b_\alpha$ in (\ref{mMajo}) should be identified
 with the mass parameters  $m_{\mathrm{IR},\alpha}^\nu$ and thus $\delta_\alpha$ measure the non-universality of $\hat b_\alpha$. As we see below, $|\delta_\alpha| \lesssim 0.1$
 are required, reflecting that the parameters $\hat b_\alpha$  have to be nearly universal.
  We analyze this issue in more detail in the next subsection.
  Alternatively,  we can require the invariance of the IR localized Lagrangian under a $\Z_2$ exchange symmetry, under which 
\be
 \hat \nu_\alpha(x,R^\prime) \leftrightarrow \nu_\alpha(x,R^\prime) \,, \;\;
 \tilde L_{1,\alpha}(x, R^\prime)  \leftrightarrow \tilde L_{2,\alpha}(x, R^\prime)\,, \;\;
 L_{\alpha}(x, R^\prime)  \leftrightarrow \hat L_{\alpha}(x,R^\prime)\,, 
\label{Z2acc}
\ee
so that we can take $m_{\mathrm{IR},\alpha}^\nu=m_{\mathrm{IR},\alpha}^l=1$ (up to an irrelevant sign per generation). We denote this constrained model as $\Z_2$-invariant model.

\subsection{Lepton Mixing}


We first discuss the phenomenological constraints on the size of $\delta_\alpha$, as defined in (\ref{delta_a}),  in the ZMA at linear order in the perturbation $\delta_\alpha$. In doing so we set
$Z_l^D=0$ and neglect the non-universality of the parameters $\rho_\alpha$, which is small for  $c_l \simeq 1/2$. Then we analyze these constraints 
numerically by taking into account the first KK level, still without considering the effect of BKT. 

We find the following analytical results,  for normally ordered neutrinos with a lightest neutrino mass $m_0=0.01$ eV and solar and atmospheric mass square differences
 $\Delta m_{\rm{sol}}^2 = 7.59 \times 10^{-5}$ eV$^2$ and 
$\Delta m_{\rm{atm}}^2 = 2.40 \times 10^{-3}$ eV$^2$ \cite{deltamsq}: in the case of $S_4$ and $A_5$, deviations from $\theta_{13}=0$ and maximal $\theta_{23}$ are proportional to the
breaking of $\mu-\tau$ symmetry $(\delta_\mu-\delta_\tau)$ and are nearly the same (the values in square brackets, if given, refer to $A_5$):
\bea
\sin \theta_{13}&  \approx  & 0.05 \, |\delta_\mu-\delta_\tau| \, ,\nn \\
 \sin^2 \theta_{23} &  \approx  & \frac{1}{2} + 0.82 \, [0.83] \,  (\delta_\mu-\delta_\tau) \, ,
 \label{S4A5} \\
 \sin^2 \theta_{12} & \approx & \frac{1}{3} \, [0.28]  +1.58 \, [1.43] \, (2\delta_e-\delta_\mu-\delta_\tau)\,. \nn
\eea
For the mixing pattern {\tt M1} arising from $\Delta(96)$ we find
\be
\begin{matrix}
\sin^2 \theta_{13} & \approx & 0.04+ 0.13 \delta_e - 0.11 \delta_\mu - 0.03 \delta_\tau\, ,\\
\label{D96M1}
 \sin^2 \theta_{23} & \approx & 0.65 -0.02\delta_e+0.76 \delta_\mu- 0.74 \, \delta_\tau\, ,\\
 \sin^2 \theta_{12} &\approx & 0.35 +3.09 \delta_e -0.67 \delta_\mu  - 2.42 \delta_\tau \, .
\end{matrix}
\ee
The results for {\tt M2} are related to these by exchanging $\delta_\mu$  and $\delta_\tau$ and by replacing the unperturbed value of $\sin^2\theta_{23}$ by $0.35$ together
with a sign change in its corrections
 (remember that {\tt M1} and {\tt M2} are related by the exchange of the second and third rows
of $V$). As can be seen, for $\delta_\mu \approx \delta_\tau$ corrections to the atmospheric mixing angle become suppressed. For the mixing pattern {\tt M3}
coming from $\Delta(384)$ we get
\be
\begin{matrix}
\sin^2 \theta_{13} & \approx & 0.01+ 0.04 \delta_e - 0.004 \delta_\mu - 0.03 \delta_\tau\, ,\\
\label{D384M3}
 \sin^2 \theta_{23} & \approx & 0.42 +0.01\delta_e+0.80 \delta_\mu- 0.81 \, \delta_\tau \, ,\\
 \sin^2 \theta_{12} &\approx & 0.34 +3.15 \delta_e -2.02 \delta_\mu  - 1.13 \delta_\tau \, .
\end{matrix} 
\ee
The mixing angles and their corrections in $\delta_\alpha$ associated with the 
patterns {\tt M3} and {\tt M4} are related in the same way as those of {\tt M1} and {\tt M2} (the unperturbed value of 
$ \sin^2 \theta_{23}$ is $0.58$ in the case of {\tt M4}). Again, $\theta_{23}$ only receives small
corrections for $\delta_\mu \approx \delta_\tau$. As (\ref{S4A5})-(\ref{D384M3}) show, the corrections to the solar mixing 
angle,  the one which is experimentally determined with best precision, are generally the largest with the coefficients of $\delta_\alpha$ being larger than one. This also implies
that the above perturbative expansion makes sense only for $|\delta_\alpha|\lesssim 0.1$.
As mentioned in \cite{S4HCHM},  the validity of the expansion in $\delta_\alpha$ strongly depends on $m_0$ and gets worse for increasing $m_0$.
In the case of inverted mass hierarchy the coefficients multiplying the linear perturbations in $\theta_{12}$  are more than one order of magnitude bigger
than the ones in (\ref{S4A5})-(\ref{D384M3}), implying that a perturbative expansion in $\delta_\alpha$ is not valid for any value of $m_0$.
This behaviour is in general expected due to the near degeneracy of the two heavier neutrinos in the case of an inversely ordered mass spectrum.
Note that in the limit of universal $\delta_\alpha$ all corrections to the mixing angles vanish.

\begin{figure}[t!]
\begin{center}
\includegraphics*[width=0.7\textwidth]{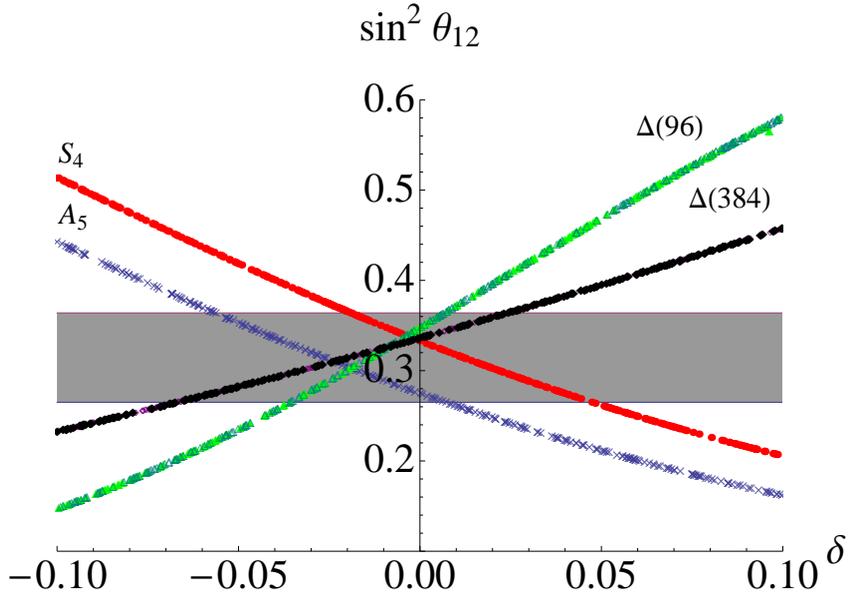}
\end{center}
\caption{\small The solar mixing angle $\theta_{12}$ as function of the deviation $\delta$, parametrizing the non-universality of the masses $m_{{\rm IR},\alpha}^\nu$, see (\ref{delta_plot}). 
The horizontal gray band corresponds to the $3 \sigma$ range as given in \cite{fogli} using 
new reactor fluxes. The different symbols and colors distinguish the various models: 
$S_4$ (red points), $A_5$ (blue crosses), $\Delta(96)$ and mixing pattern {\tt M1} (light green full triangles), $\Delta(96)$ and mixing pattern {\tt M2} (dark green open triangles) , $\Delta(384)$ 
and mixing pattern {\tt M3} (black full diamonds) and $\Delta(384)$ and mixing pattern {\tt M4} (violet open diamonds).  The curves for {\tt M1} and {\tt M2} and {\tt M3} and {\tt M4}, respectively,
lie on top of each other due to the parametrization chosen in (\ref{delta_plot}).
The plots refer to $c_l=0.52$, $c_\nu = -0.365$, $h=1/3$ and normal neutrino mass hierarchy. 
The masses $m_{\mathrm{UV},\alpha}$ are chosen such that  
 the lightest neutrino mass is $m_0=0.01$ eV and the values of the solar and atmospheric mass square differences $\Delta m_{\rm{sol}}^2 = 7.59 \times 10^{-5}$ eV$^2$ and 
$\Delta m_{\rm{atm}}^2 = 2.40 \times 10^{-3}$ eV$^2$ \cite{deltamsq} are reproduced using the ZMA. For simplicity, we take the IR mass terms $m_{{\rm IR},\alpha}^l=1$ and set all BKT to zero. 
 We vary  $m_{{\rm IR},0}^\nu$, see (\ref{delta_a}),  between $0.3$ and $1$.}
\label{fig:theta12}
\end{figure}

In our numerical analysis we discuss the maximal allowed size of the corrections $\delta_\alpha$ in order to keep accordance with experimental data.
We choose a particular parametrization for the deviations of $m_{\mathrm{IR},\alpha}^\nu$ from universality in terms of only one parameter $\delta$: 
\bea
S_4 \, , \, A_5 &:& \delta_e=0, \delta_\mu=\delta, \delta_\tau=0 \, ,\nn\\
\label{delta_plot}
\Delta(96), \mbox{{\tt M1}} \, , \,  \Delta(384), \mbox{{\tt M3}} &:& \delta_e=\delta, \delta_\mu=\delta, \delta_\tau=0 \, , \\
\Delta(96), \mbox{{\tt M2}} \, , \,  \Delta(384), \mbox{{\tt M4}} &:& \delta_e=\delta, \delta_\mu=0, \delta_\tau=\delta \, . \nn
\eea
This particular parametrization leads to the same results for the mixing angles $\theta_{13}$ and $\theta_{12}$ in case of the patterns {\tt M1} and {\tt M2} as well as for
{\tt M3} and {\tt M4}. The atmospheric mixing angle acquires at the same time a correction which is the same in size, but opposite in sign for {\tt M1} ({\tt M3}) and {\tt M2}
({\tt M4}). As already obvious from the analytical results, the corrections are the largest for the angle $\theta_{12}$ whose dependence on $\delta$ we report in 
figure \ref{fig:theta12}, together with its experimentally allowed $3\sigma$ range using the new estimate of reactor anti-neutrino fluxes \cite{fogli}. Clearly, $|\delta|$ has to be smaller than $0.07$ in all
cases. Depending on the unperturbed value of $\sin^2\theta_{12}$, either positive or negative $\delta$ is better compatible with the data for increasing $|\delta|$. All cases apart from $X=S_4$ prefer $\delta<0$
for the particular choice (\ref{delta_plot}).
We do not show our results for the other two mixing angles, since in the case of $\theta_{13}$ all corrections for $|\delta| \lesssim 0.1$ are small: 
in the case of $S_4$ and $A_5$ $\sin^2\theta_{13} \lesssim 4 \times 10^{-5}$ holds, compare (\ref{S4A5}), while
 for patterns {\tt M1} and {\tt M2} we find $0.041 \lesssim \sin^2 \theta_{13} \lesssim 0.047$
and for {\tt M3} and {\tt M4} we get $0.0084 \lesssim \sin^2 \theta_{13} \lesssim 0.015$.
We clearly see that corrections associated with the non-universality of the neutrino Dirac mass terms are not sufficient in the cases $X=S_4$ and $X=A_5$ to explain $\theta_{13} \sim 0.1 \div 0.2$.
For $|\delta| \lesssim 0.1$,  the corrections to the atmospheric mixing angle always keep $\sin^2\theta_{23}$ in its
$3\sigma$ range \cite{fogli} for $S_4$, $A_5$ and $\Delta(384)$. Since the unperturbed value of $\sin^2\theta_{23}$ is at the edge of the $3\sigma$ range for the patterns {\tt M1}, 
$\sin^2\theta_{23} \approx 0.65$, and {\tt M2}, $\sin^2 \theta_{23} \approx 0.35$, 
corrections with $\delta<0$ for the parametrization (\ref{delta_plot}) are welcome because they improve the agreement with the results from global fits. 
At the same time $\theta_{12}$, see figure \ref{fig:theta12}, remains in its
experimentally allowed $3 \sigma$ range for negative $\delta$ with $|\delta| \lesssim 0.04$. Generally speaking, the patterns {\tt M3} and {\tt M4} are the most promising ones even taking
into account corrections coming from the non-universality of the neutrino Dirac mass terms.\footnote{There is a slight dependence of which of the two patterns performs best on the used global fit analysis:
using \cite{fogli,maltoni} we find pattern {\tt M3} to be the best one, while \cite{schwetz} prefers {\tt M4} over {\tt M3}. This difference originates from the fact that in \cite{schwetz} the best fit value of 
$\sin^2 \theta_{23}$ is larger than 0.5, whereas it is smaller than 0.5 in \cite{fogli,maltoni}.}
 In the case of the patterns {\tt M1} and {\tt M2} these corrections help to improve the
accordance with the experimental data; however, these patterns are not favoured by the latter. $S_4$ and $A_5$ mainly fail to give a good fit to the data because of the too small value of $\theta_{13}$.

The effects of BKT and of non-universal $\rho$ on the lepton mixing, neglected in the above study being sub-leading, become relevant
in the $\Z_2$-invariant model, where $\delta_\alpha$ vanish. In the latter model we find numerically for $S_4$ and for $A_5$ (see caption of figure 
\ref{fig:mue-gamma} for details on the chosen parameters)
\be
0.32 \lesssim \sin^2 \theta_{12} \lesssim 0.35
\;\;\; \mbox{and} \;\;\;
0.27 \lesssim \sin^2 \theta_{12} \lesssim 0.29 \, ,
\ee
respectively,  and
\be
0.48 \lesssim \sin^2 \theta_{23} \lesssim 0.52 \, ,
\ee
showing that $\sin^2 \theta_{12,23}$ only get corrected by less than $0.03$. Corrections to $\theta_{13}$ are negligible and thus
 $\theta_{13} \sim 0.1 \div 0.2$ \cite{fogli,schwetz,maltoni} cannot be achieved in models with $S_4$ and $A_5$,  by taking into account 
the BKT or the deviation of the parameters $\rho$ from universality.
For $\Delta(96)$, {\tt M1} [{\tt M2}] we get
\be
0.63 [0.32] \lesssim \sin^2 \theta_{23} \lesssim 0.69 [0.36] \, ,
\ee
keeping the atmospheric mixing angle compatible with data only at the $3 \sigma$ level \cite{fogli},
as well as
\be
0.34 \lesssim \sin^2 \theta_{12} \lesssim 0.37 \;\;\; \mbox{and} \;\;\;
0.04 \lesssim \sin^2 \theta_{13} \lesssim 0.053 [0.047] \, .
\ee
For $\Delta(384)$, {\tt M3} [{\tt M4}] we find analogously
\be
0.40 [0.56] \lesssim \sin^2 \theta_{23} \lesssim 0.44 [0.59] \, ,
\ee
respectively, and
\be
0.32 \lesssim \sin^2 \theta_{12} \lesssim 0.35 \;\;\; \mbox{and} \;\;\;
0.01 \lesssim \sin^2 \theta_{13} \lesssim 0.012  \, .
\ee

\subsection{LFV Processes}

\label{subsec:Fbounds}

In this subsection we examine the bounds coming from LFV processes in all six models proposed (see \cite{Agashe:2006iy} for early analysis of LFV bounds in warped models).
As discussed in detail in \cite{S4HCHM} and repeated in section 2, these processes are non-negligible only when
the leading flavour violating BKT (\ref{LBKT}) at the UV brane are taken into account.
The tree-level decay $\mu\rightarrow 3e$, $\mu-e$ conversion in nuclei and $\mu\to e\gamma$ depend quadratically on the off-diagonal entry 
$(e\mu)$ of the matrix (\ref{BKTeff2}). The relevant combination of $\hat z_{\alpha l}$ varies
from case to case because $Z_l$ depend on the mixing matrix $V$. 
\begin{figure}[t!]
\begin{center}
\includegraphics*[width=0.7\textwidth]{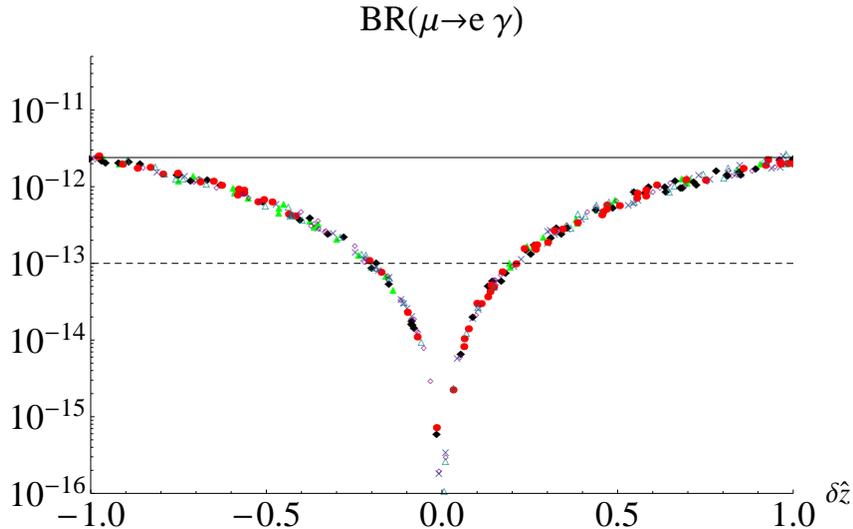}
\end{center}
\caption{\small Branching ratio of $\mu\rightarrow e \gamma$ as a function of the UV BKT $\delta \hat z = 3(\hat Z_l)_{e\mu}$ in the $\Z_2$-invariant model. 
The continuous and dashed lines are the current ($BR(\mu\to e\gamma) < 2.4 \times 10^{-12}$) and the expected future bound ($BR(\mu\to e\gamma) < 10^{-13}$) given by the 
MEG Collaboration \cite{MEG}, respectively.  The parameters
$c_l$, $c_\nu$, $h$, $m_0$ and the neutrino mass hierarchy as well as the color coding are chosen as in figure \ref{fig:theta12}. Note that no constraints
coming from lepton mixing are taken into account in this data set; however, the constraints coming from gauge coupling deviations are satisfied by each point plotted.}
\label{fig:mue-gamma}
\end{figure}
We plot in figure \ref{fig:mue-gamma} the bounds arising from $\mu\rightarrow e \gamma$ 
as a function of $\delta \hat z = 3(\hat Z_l)_{e\mu}$ for the $\Z_2$-invariant model,\footnote{The factor 3 in the definition naturally arises in the $S_4$ case, and is left
to match the convention used in \cite{S4HCHM}.} obtained by a numerical computation in which the first KK mode of each tower of states has been kept.
$BR(\mu\rightarrow e \gamma)$ is essentially the same in all models and shows the expected quadratic dependence on $\delta \hat z$. 
The branching ratio is always below the current experimental limit for $|\delta \hat z|\lesssim 1$ and the expected future MEG bound $BR(\mu\to e\gamma) < 10^{-13}$
will require $|\delta \hat z| \lesssim 0.2$. Similar results hold for the model with no $\Z_2$ exchange symmetry since the IR masses $m_{\rm{IR},\alpha}^{l,\nu}$ do not play an important role.
Finally we note that  in all models the branching ratios of $\mu\to 3 e$, $\mu-e$ conversion in Ti as well as of the radiative $\tau$ lepton decays are below their experimental bounds for $|\delta \hat z| \lesssim 1$.

\subsection{EDMs}

\label{subsec:edmbounds}

Lepton EDMs in the 5D model are completely calculable, because gauge invariance forbids the appearance of uncalculable 5D bulk or boundary operators
that reduce, upon KK reduction in  4D,
to the EDM operators $(-i/2) \bar l_\alpha\sigma^{\mu\nu} \gamma_5 l_\alpha F_{\mu\nu}$. The size of $d_\alpha$ is negligible, due to the relatively few sources of complex parameters.
 The bulk parameters $c_l$, $c_\alpha$ and $c_\nu$ are real, while
the IR and UV mass terms contain in total nine complex phases. Those of the IR mass terms
$m_{{\rm IR},\alpha}^{l,\nu}=|m_{{\rm IR},\alpha}^{l,\nu}|e^{i\theta_\alpha^{l,\nu}}$ can be removed by a simple field redefinition:
\be
\xi_{\nu,\alpha}\rightarrow e^{-i\theta_\alpha^\nu} \xi_{\nu,\alpha}\,, \ \ \ \ 
\xi_{e,\alpha}\rightarrow e^{-i\theta_\alpha^l} \xi_{e,\alpha}\,.
\label{edmrota}
\ee
Note that the terms in the bulk Lagrangian do not change if the transformations (\ref{edmrota}) are applied, because the terms involving the 5D fields $\xi_{l,\alpha}$
and $\xi_{\nu,\alpha}$ are invariant under the flavour symmetry $U(3)$ and those containing $\xi_{e,\alpha}$ under $U(1)^3$.
In the field basis of (\ref{edmrota}), all phases are encoded in the UV Majorana mass terms: the phases associated with the UV mass terms $m_{{\rm UV},\alpha}$
as well as the phases  $\theta_\alpha^\nu$. As has been shown in \cite{S4HCHM} (cf. (4.34)) by considering the KK expansion of the RH neutrinos $\nu_{\alpha R}$, the UV localized Majorana mass term 
gives effectively rise to only one heavy RH neutrino per generation, with mass of order $10^{12\div 13}$ GeV, while the remaining orthonormal combinations of KK states are not sensitive to the UV Majorana mass term
and have masses setting in at a few TeV.  As consequence, the EDMs induced by the UV Majorana mass terms are mediated by these three heavy states 
and are completely negligible.
Notice that the transformations (\ref{edmrota}) do not involve the fields $\xi_{l,\alpha}$ and hence no phases appear in the BKT (\ref{LBKT}). 
The next-to-leading flavour violating BKT at the UV brane is $Z_\nu$ in (\ref{Z4Dgen}), which acquire non-trivial phases
after the transformations (\ref{edmrota}). Due to the field localization of $\nu_{\alpha R}$ in the extra dimension, however, the effective BKT are strongly suppressed, $Z_\nu\lesssim 10^{-5}$, 
if values of $c_\nu$ are used which are suitable
for reproducing correctly the scale of light neutrino masses, e.g. $c_\nu=-0.365$ like in figures \ref{fig:theta12} and \ref{fig:mue-gamma} (this is equivalent to the suppression mechanism
of $Z_\nu$ explained below (\ref{Z4Dgen}) for a composite $\nu_R^\alpha$). 
Using the estimate (\ref{dledm2}) with $Z_l$ being replaced by $Z_\nu$, we find that lepton EDMs are well below the current experimental bounds.

 \section{Conclusions} 

We have extended the class of 4D HCHM based on the non-abelian flavour group $S_4\times \Z_3$ introduced in \cite{S4HCHM} to general non-abelian discrete groups
of the form $X \times \Z_N$.
This allows to consider HCHM which predict a promising lepton mixing pattern with non-vanishing $\theta_{13} \sim 0.1 \div 0.2$ \cite{dATFH11} as favoured by current data.
In a 5D scenario with Majorana neutrinos, we have computed in detail the lepton mixing for four particular choices of $X \times \Z_N$.
We have shown that flavour symmetry breaking effects at the IR brane affecting neutrinos 
can be at most $10 \%$ in order to not perturb too much the original predictions for the mixing angles. We have argued, like in \cite{S4HCHM}, that a $\Z_2$ exchange symmetry
can be imposed on the IR brane avoiding these restrictions. All LFV processes and EDMs are below the experimental bounds.

As discussed in \cite{S4HCHM}, the typical mass scale of the vector-like leptonic fermion resonances in the 5D models is around 2 TeV. The prospects to produce and observe these resonances at the LHC are unfortunately quite limited. A possible signature might be the observation of the decay
$\mu\rightarrow e\gamma$, considering that its typical branching ratio
is within the range of the expected future bound of  the MEG experiment.

It would be interesting to extend our considerations to the quark sector and discuss quark mixing in a similar manner.

\section*{Acknowledgments}

M.S. would like to thank Andrea Romanino  for useful discussions  and Kaustubh Agashe for a useful e-mail correspondence.
C.H. would like to thank Ferruccio Feruglio for useful discussions and the Aspen Center for Physics for kind hospitality during the preparation of this work.
  
\appendix

\section{Group Theory of $S_4$, $A_5$, $\Delta(96)$ and $\Delta(384)$}
\label{appendix}

In this appendix we report some details of the group theory of $S_4$, $A_5$, $\Delta(96)$ and $\Delta(384)$. 
All symmetries can be defined in terms of two generators $S$ and $T$. We show the relations the latter have to fulfill
in order to generate one of the groups, $S_4$, $A_5$, $\Delta(96)$ and $\Delta(384)$, and give their explicit realization
for an irreducible triplet. The basis chosen for $S$ and $T$ is such that the Lagrangians  introduced
in (\ref{Lele}), (\ref{IRmasses}) and  (\ref{UVmasses})  are reproduced.  In the following we report the explicit form of $S$. The
generator $T$ can be easily computed using (\ref{GeDiag}) and table \ref{table:groups}, once the product of $S$ and $T$ giving $G_N$ is known.

In the case of $S_4$, $S$ and $T$ fulfill the relations \cite{S4_BM}
\begin{equation} 
S^2 =  {1} \;\; , \;\;\; T^4=  {1} \;\; , \;\;\; (S T)^3 =  {1} \;\;.
\end{equation}
The explicit form of $S$  can be chosen as
\begin{equation}
S =\frac{1}{3} \, \left( \begin{array}{ccc}
1 & -2 & -2\\
-2 & -2 & 1\\
-2 & 1 & -2
\end{array}
\right) 
\; .
\end{equation}
The generators of $G_e$ and $G_\nu$ are the following
\begin{equation}
G_N = (S T)^2 \,, \ \ \  G_1 = S\,, \ \ \ G_2 = (S T^2)^2  \; .
\end{equation}
The explicit form of $T$ is then
\begin{equation}
T = S G_N^2 = 
\frac{1}{3} \, \left( \begin{array}{ccc}
1 & -2 \, \omega_3 & -2 \, \omega_3^2\\
-2 & -2 \, \omega_3 & \omega_3^2\\
-2 & \omega_3 & -2 \, \omega_3^2
\end{array}
\right)  \; .
\end{equation}
We also briefly comment on the relation between the generators $S$ and $T$ used here and the set of generators $\tilde{S}$, $\tilde{T}$ and $U$ used in
\cite{S4HCHM} in order to describe the group $S_4$. One can check that $\tilde{T}$
is similar to $G_N=(S T)^2$, $\tilde{S}$ to $G_2 =(S T^2)^2$ and $U$ to $G_1 G_2=T^2 S T^2$.

The group $A_5$ is generated through $S$ and $T$ being subject to the conditions \cite{A5FP}
\begin{equation}
S^2 =  {1} \,\; , \;\;\; T^5=  {1} \;\; , \;\;\; (S T)^3 =  {1} \;\;.
\end{equation}
We choose the realization of $S$ to be 
\begin{equation}
S =\frac{1}{\sqrt{5}} \, \left( \begin{array}{ccc}
1 & \sqrt{2} & \sqrt{2}\\
\sqrt{2} & -\phi & \frac{1}{\phi}\\
\sqrt{2} & \frac{1}{\phi} & -\phi
\end{array}
\right) 
\end{equation}
with $\phi=(1+\sqrt{5})/2$  for one of the irreducible triplets.
In this case the remnant subgroups $G_e$ and $G_\nu$ are generated through
\begin{equation}
G_N =  T\,, \ \ \ G_1 = S\,, \ \ \ G_2 =  T^2 S T^3 S T^2 \; .
\end{equation}

The group $\Delta(96)$ is generated through $S$ and $T$ fulfilling the relations \cite{dATFH11}
\begin{equation}
S^2 =  {1} \,\; , \;\;\; T^8=  {1} \;\; , \;\;\; (S T)^3 =  {1} \;\; , \;\;\; (S T^{-1} S T)^3 =   {1} \; .
\end{equation}
In order to realize the mixing pattern called \texttt{M1}, the most convenient choice of basis for $S$ for a faithful
irreducible triplet is
\begin{equation}
S =\frac{1}{3}\, \left( \begin{array}{ccc}
-1+\sqrt{3} & -1 & -1-\sqrt{3}\\
-1 & -1-\sqrt{3} & -1+\sqrt{3}\\
-1-\sqrt{3} & -1+\sqrt{3} & -1
\end{array}
\right) 
\; .
\label{SM1}
\end{equation}
For the other mixing pattern \texttt{M2} it is convenient to choose  $S$ like in (\ref{SM1}), however with second and third rows and columns
exchanged, respectively. This is clear because the mixing matrices $V$ associated with {\tt M1} and {\tt M2} are related by the exchange of the
second and third rows.
Independently of the mixing pattern, the remnant subgroups $G_e$ and $G_\nu$ are generated through \cite{dATFH11}
\begin{equation}
G_N = S T \,, \ \ \  G_1 = S\,, \ \ \ G_2 = (S T^4)^2  \,.
\end{equation}

Similarly, the group $\Delta(384)$ is generated with $S$ and $T$ fulfilling \cite{dATFH11}
\begin{equation}
S^2 =  {1} \,\; , \;\;\; T^{16}=  {1} \;\; , \;\;\; (S T)^3 =  {1} \;\; , \;\;\; (S T^{-1} S T)^3 =   {1} \; .
\end{equation}
Again, it is convenient to choose two different bases for a faithful irreducible triplet representation in order to generate the mixing patterns $\texttt{M3}$ and $\texttt{M4}$, respectively. For
a model incorporating the pattern $\texttt{M3}$ we choose
\begin{equation}
S= \frac{1}{6} \, \left( \begin{array}{ccc}
 -2 +\sqrt{2}+\sqrt{6} & -2 (1+\sqrt{2}) & -2+\sqrt{2} -\sqrt{6}\\
 -2(1+\sqrt{2}) & -2+\sqrt{2}-\sqrt{6} & 2(-1+\sqrt{2+\sqrt{3}})\\
-2+\sqrt{2}-\sqrt{6} & 2(-1+\sqrt{2+\sqrt{3}}) & -2(1+\sqrt{2})
\end{array}
\right)  \; .
\label{SM3}
\end{equation}
For an explicit model leading to the pattern \texttt{M4}, the basis in which $S$ is like in (\ref{SM3}) with second and third rows
and columns exchanged, respectively, is the most appropriate one. Again, this is obvious considering the relation of the mixing patterns {\tt M3} and {\tt M4}.
Independently of the mixing pattern, the remnant subgroups $G_e$ and $G_\nu$ are generated through \cite{dATFH11}
\begin{equation}
G_N = S T \,, \ \ \ G_1 = S\,, \ \ \ G_2 = (S T^8)^2  \, .
\end{equation}


\end{document}